\documentclass[pdflatex,sn-mathphys-num]{sn-jnl}

\usepackage{graphicx}%
\usepackage{multirow}%
\usepackage{amsmath,amssymb,amsfonts}%
\usepackage{amsthm}%
\usepackage{mathrsfs}%
\usepackage[title]{appendix}%
\usepackage{xcolor}%
\usepackage{textcomp}%
\usepackage{manyfoot}%
\usepackage{booktabs}%
\usepackage{algorithm}%
\usepackage{algorithmicx}%
\usepackage{algpseudocode}%
\usepackage{listings}%

\theoremstyle{thmstyleone}%
\theoremstyle{thmstyletwo}%

\theoremstyle{thmstylethree}%

\renewcommand{\figurename}{Figure}

\begin{document}

\title[Article Title]{Electronic structures and multi-orbital models of La$_3$Ni$_2$O$_7$ thin films at ambient pressure}


\author[1,2]{\fnm{Xunwu} \sur{Hu}}
\equalcont{These authors contributed equally to this work.}

\author[2]{\fnm{Wenyuan} \sur{Qiu}}
\equalcont{These authors contributed equally to this work.}

\author[2]{\fnm{Cui-Qun} \sur{Chen}}

\author[2]{\fnm{Zhihui} \sur{Luo}}

\author*[2]{\fnm{Dao-Xin} \sur{Yao}}\email{yaodaox@mail.sysu.edu.cn}

\affil[1]{\orgdiv{Department of Physics, College of Physics and Optoelectronic Engineering}, \orgname{Jinan University}, \city{Guangzhou}, \postcode{510632}, \country{China}}

\affil[2]{\orgdiv{Center for Neutron Science and Technology, Guangdong Provincial Key Laboratory of Magnetoelectric Physics and Devices, State Key Laboratory of Optoelectronic Materials and Technologies, School of Physics}, \orgname{Sun Yat-Sen University}, \city{Guangzhou}, \postcode{510275}, \country{China}}


\abstract{The recent discovery of superconductivity with a transition temperature $T_c$\ exceeding 40 K in La$_3$Ni$_2$O$_7$ and (La,Pr)$_{3}$Ni$_2$O$_7$ thin films at ambient pressure marks a significant breakthrough in the field of nickelate superconductors. Using density functional theory (DFT), we propose a double-stacked two-orbital effective model for La$_3$Ni$_2$O$_7$ thin film based on the Ni$-e_g$ orbitals. Our analysis of the Fermi surface reveals three electron pockets ($\alpha,\alpha^{\prime},\beta$) and two hole pockets ($\gamma,\gamma^{\prime}$), where the additional $\alpha^{\prime}$ and $\gamma^{\prime}$ pockets arise from inter-stack interactions. Furthermore, we introduce a high-energy model that incorporates O$-p$ orbitals to facilitate future studies. Calculations of spin susceptibility within the random phase approximation (RPA) indicate that magnetic correlations are enhanced by nesting of the $\gamma$ pocket, which is predominantly derived from the Ni$-d_{z^2}$ orbital. Our results provide a theoretical foundation for understanding the electronic and magnetic properties of La$_3$Ni$_2$O$_7$ thin films. 
}

\maketitle

\section{Introduction}\label{sec1}
The discovery of superconductivity in the Ruddlesden-Popper (RP) bilayer nickelate La$_3$Ni$_2$O$_7$ at a transition temperature $T_c$ near 80 K under high pressure ($\sim$ 14 Gpa) has generated significant interest in the field of unconventional superconductivity~\cite{sun2023signatures}. The subsequent observation of superconductivity in the trilayer nickelate La$_4$Ni$_3$O$_{10}$ under similar conditions further underscores the universality of superconductivity in nickelates~\cite{zhu_superconductivity_2024}. These discoveries have motivated extensive theoretical~\cite{luo2023bilayer, zhang2023electronic,lechermann2023electronic,luo2024high, wu2024superexchange,  cpl_41_7_077402,   shilenko2023correlated, yang2023interlayer, zhang2023trends,christiansson2023correlated,shen2023effective, oh2023type, liu2023s,liao2023electron,yang2023possible, PhysRevB.110.014503, PhysRevB.111.075140,PhysRevB.110.235155,kaneko2024pair,ouyang2024absence,PhysRevB.109.144511, heier2024competing, zhang2024structural, zhang2024electronic,tian2024correlation, ryee2024quenched,zhang2024strong, ni_spin_2025, lu2024interlayer, qu2024bilayer, yang2024strong, fan2024superconductivity,   sakakibara2024possible, cao2024flat, jiang2024pressure,chen2025charge} and experimental~\cite{yang_orbital-dependent_2024, NPzhang,Hou_2023, PhysRevB.110.134520, wang_pressure-induced_2024, liu_electronic_2024, li_signature_2024, li_identification_2025,liu2024electronic} investigations into the microscopic mechanisms of unconventional superconductivity. However, the requirement of high pressure for superconductivity in La$_3$Ni$_2$O$_7$ and La$_4$Ni$_3$O$_{10}$ presents significant experimental challenges, which limit in-depth investigations. This has driven efforts to stabilize superconductivity under ambient pressure, facilitating a more comprehensive understanding of its underlying mechanisms.

Recent studies have reported superconductivity with $T_c$ exceeding 40 K in La$_3$Ni$_2$O$_7$~\cite{ko_signatures_2025}, (La,Pr)$_{3}$Ni$_2$O$_7$~\cite{zhou_ambient-pressure_2025,liu2025superconductivitynormalstatetransportcompressively} thin films at ambient pressure, marking a significant breakthrough in nickelate superconductors. X-ray absorption spectroscopy (XAS) reveals that the Ni ions in La$_3$Ni$_2$O$_7$ thin film retain a mixed valence state similar to that of the bulk form~\cite{ko_signatures_2025}. Furthermore, scanning transmission electron microscopy (STEM) confirms that their microscopic structure closely resembles that of the bulk phase of high pressure, adopting a tetragonal crystal structure~\cite{zhou_ambient-pressure_2025}. In particular, the $T_c$ in the thin film is tunable via the in-plane lattice constant but remains relatively insensitive to the out-of-plane parameter. DFT calculation and angle-resolved photoemission spectroscopy (ARPES) measurement~\cite{li2025photoemissionevidencemultiorbitalholedoping,shen2025nodelesssuperconductinggapelectronboson} suggest that Ni-$d_{x^2-y^2}$ and Ni-$d_{z^2}$ orbitals contribute states near the Fermi level, highlighting the similarity between the thin film and the bulk system in terms of electronic structure and superconducting properties. Despite these advances, most theoretical studies have relied on simplified half-unit-cell (Half-UC) slab model~\cite{yue2025,shao2025}, which may not fully capture the effects of dimensionality on the electronic structure. A more comprehensive approach incorporating the full-unit-cell slab model is therefore needed to provide a more accurate description of this system.

In this paper, we employ slab models of La$_3$Ni$_2$O$_7$ thin films to investigate their electronic structures in various thicknesses, including three-unit-cell (Three-UC), one-unit-cell (One-UC), and Half-UC configurations. Except for the Half-UC case, each slab model retains a complete unit cell (UC), with the two bilayers denoted as Stack 1 and Stack 2, enabling a systematic exploration of the relationship between dimensionality and electronic properties. Using first-principles calculations, we propose a double-stacked two-orbital effective model for the One-UC slab structure, based on Ni$-e_g$ orbitals. Our analysis reveals the presence of three electron pockets ($\alpha,\alpha^{\prime},\beta$) and two hole pockets ($\gamma,\gamma^{\prime}$) on the Fermi surface, where inter-stack interactions give rise to the additional $\alpha^{\prime}$ and $\gamma^{\prime}$ pockets. Furthermore, we extend our model by incorporating O$-p$ orbitals into a high-energy framework to facilitate future studies. Spin susceptibility calculations within the RPA indicate pronounced magnetic correlations, primarily driven by nesting effects of the $\gamma$ pocket, which is predominantly derived from the Ni$-d_{z^2}$ orbital. Our results provide theoretical framework for understanding the interplay among the dimensionality, magnetism, and superconductivity in La$_3$Ni$_2$O$_7$ thin films, offering key insights for future theoretical and experimental investigations.

\section{Results}\label{sec2}

\subsection{Slab structures}
\begin{figure}
        \centering
	\includegraphics[scale=0.45]{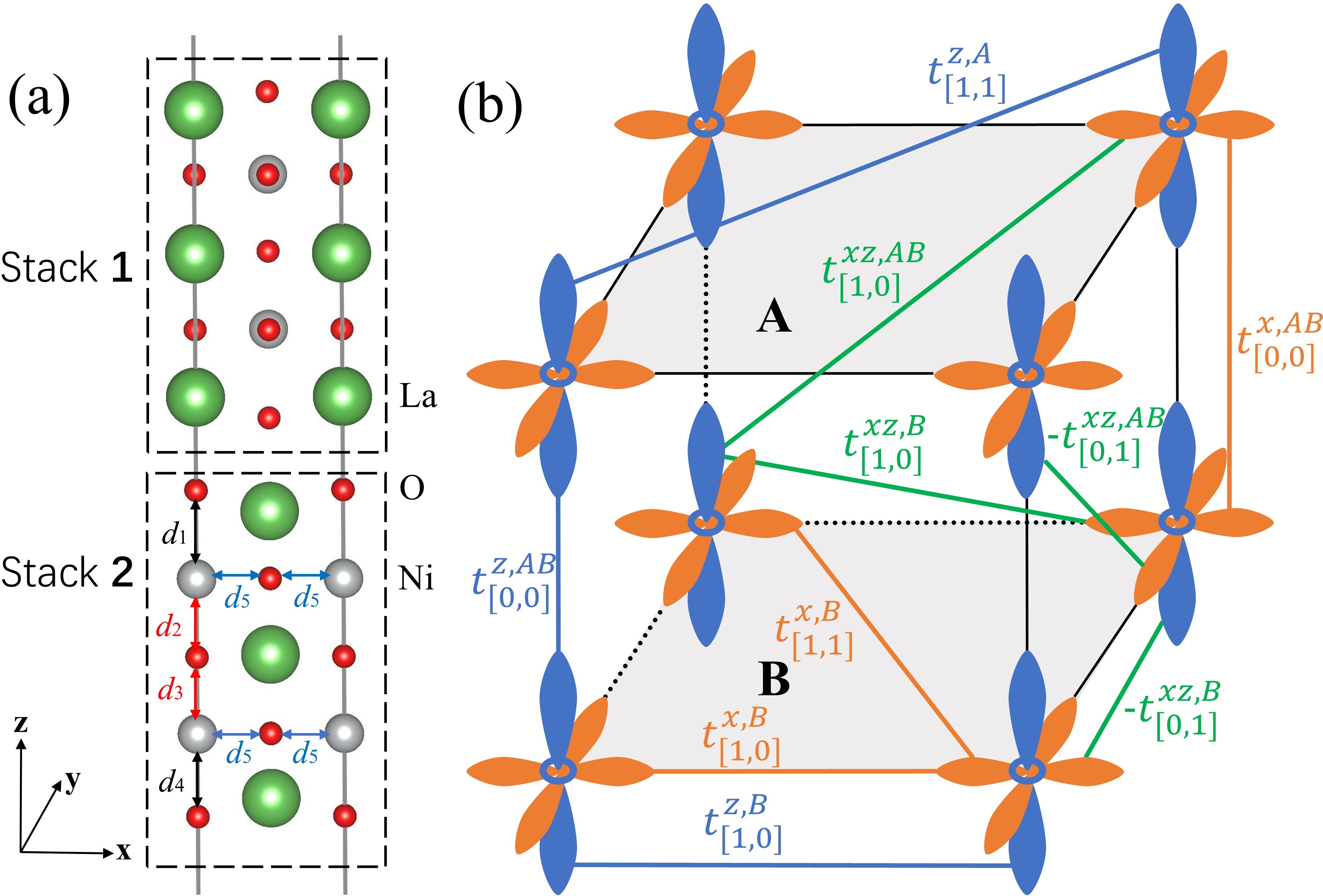}\caption{\label{fig1}
    Slab structure and schematic illustration of the hopping parameters used in the La$_3$Ni$_2$O$_7$ thin-film model.
    (a) The free-standing One-UC slab structure of La$_3$Ni$_2$O$_7$, where the two bilayers are labeled as Stack 1 and Stack 2. Green, gray, and red spheres denote La, Ni, and O atoms, respectively. The outer-apical~($d_1$ and $d_4$), inner-apical ($d_2$ and $d_3$), and in-plane ($d_5$) Ni-O bond lengths are indicated by black, red, and blue arrows, respectively. (b) Schematic illustration of the hopping parameters in the La$_3$Ni$_2$O$_7$ thin films, highlighting the Ni$-d_{x^2-y^2}$~(orange) and $d_{z^2}$~(blue) orbitals. Only nearest-neighbor hopping terms are shown.}
\end{figure}

We begin our investigation of the La$_3$Ni$_2$O$_7$ thin films by modeling the slab structures using DFT calculations. Bulk La$_3$Ni$_2$O$_7$ belongs to the $n=2$ Ruddlesden-Popper phase, characterized by a UC comprising two corner-sharing NiO$_6$ octahedra bilayers stacking along the $c$ axis~\cite{sun2023signatures}. In its thin-film form, La$_3$Ni$_2$O$_7$ adopts a tetragonal structure similar to the high-pressure bulk phase, distinguished by an apical Ni-O-Ni bond angle approaching 180$^{\circ}$~\cite{ko_signatures_2025,bhatt2025resolvingstructuraloriginssuperconductivity}. To systematically explore the structural and electronic properties of La$_3$Ni$_2$O$_7$ thin films, we consider slab structures of varying thicknesses along the out-of-plane direction, including Three-UC, One-UC, and Half-UC slabs. To minimize interactions between preiodic images, a vacuum spacing exceeding 16 $\AA$ is introduced along the $c$ axis. 

The One-UC slab consists of two stacked NiO$_6$ octahedra bilayers, which can adopt two distinct stacking configurations, Stack 1 and Stack 2, as illustrated in Figure~\ref{fig1}(a). In all slab models, we consider a free-standing geometry with a vacuum layer to eliminate inter-slab interactions. The effect of epitaxial strain from the substrate is incorporated by fixing the in-plane lattice constant to the experimentally measured value of $a = 3.77 \AA $~\cite{ko_signatures_2025}. Although this approach does not capture all aspects of the experimental system—particularly in thicker films, where subtle differences between the top and bottom layers may occur—it provides a reasonable and reliable approximation for the few-layer systems considered in this study. The Ni-O bond lengths, labeled as $d_1$, $d_2$, $d_3$, $d_4$, and $d_5$ in Figure~\ref{fig1}(a), are summarized in Table \ref{tab1} for different slab structures and effective Hubbard parameters ($U_{eff} = 0$ eV and $U_{eff} = 2$ eV). For clarity, only the bond distances of the middle UC are reported for the Three-UC slab structure.
\begin{table}[t]
	\caption{\label{tab1} The Ni-O bond lengths ($d$) in NiO$_6$ octahedra of La$_{3}$Ni$_{2}$O$_{7}$ slab structures under $U_{eff} = 0$ eV and $U_{eff} = 2$ eV. The unit of $d$ are in $\AA$.}
		\centering
		\begin{tabular}{cccccc}
            \toprule
			\multicolumn{1}{c}{$U_{eff}$ (eV)} & Stack & $d$  & Three-UC & One-UC  & Half-UC  \\
			\hline
			\multirow{4}{*}{$U_{eff} = 0$}
			&  1 & $d_1$  &  2.292 &  2.120 & 2.116 \\ 
			&    & $d_2$  &  1.965 &  1.973 & 2.003 \\
			&    & $d_3$  &  1.965 &  2.003 & 2.003 \\
			&    & $d_4$  &  2.282 &  2.276 & 2.116 \\
                &    & $d_5$  &  1.885 &  1.885 & 1.885 \\
			\cmidrule{2-6}
			& 2  & $d_1$  &  2.282 &  2.276 & -- \\
			&    & $d_2$  &  1.965 &  2.003 & -- \\
			&    & $d_3$  &  1.965 &  1.973 & -- \\
			&    & $d_4$  &  2.292 &  2.120 & -- \\
			&    & $d_5$  &  1.885 &  1.885 & -- \\
			\hline
			\multirow{4}{*}{$U_{eff} = 2$} 
			& 1  & $d_1$  &  2.308 &  2.133 & 2.120 \\
			&    & $d_2$  &  1.960 &  1.944 & 1.996 \\
			&    & $d_3$  &  1.960 &  2.012 & 1.996 \\
			&    & $d_4$  &  2.307 &  2.364 & 2.120 \\
			&    & $d_5$  &  1.885 &  1.885 & 1.885 \\
			\cmidrule{2-6}
			& 2  & $d_1$  &  2.307 &  2.364 & -- \\
			&    & $d_2$  &  1.960 &  2.012 & -- \\
			&    & $d_3$  &  1.960 &  1.944 & -- \\
			&    & $d_4$  &  2.308 &  2.133 & -- \\
			&    & $d_5$  &  1.885 &  1.885 & -- \\
            \botrule
		\end{tabular}
\end{table}

For $U_{eff} = 0$~eV, the in-plane Ni-O bond ($d_5$) is the shortest, while the outer-apical Ni-O bonds ($d_1$ and $d_4$) are the longest, consistent with the high-pressure phase of La$_3$Ni$_2$O$_7$. Different slab structures yield nearly identical results in terms of bond lengths. In the Three-UC and One-UC slabs, $d_1$ and $d_4$ vary across stacking configurations, while $d_2$ and $d_3$ also exhibit stack-dependent variations, indicating symmetry breaking between different stacks. The Ni-O bond lengths in these slabs follow a specific symmetry relation: in Stack 1, $d_1$, $d_2$, $d_3$, $d_4$ correspond to $d_4$, $d_3$, $d_2$, $d_1$ in Stack 2. In contrast, the Half-UC slab exhibits a distinct structural behavior, characterized by Ni-O bond lengths that are symmetric about the central oxygen atom, differing from the asymmetric distortions observed in the other slab structures. This difference arises from the underlying structural symmetry. In Half-UC, the slab posesses $m_z$ symmetry, with the mirror plane located between the two NiO layers. This symmetry enforces bond length equivalence with respect to the mirror plane, resulting in two nonequivalent vertical bond lengths. However, the One-UC and Three-UC slabs exhibit a different symmetry operation, $\{m_z|\frac{1}{2},\frac{1}{2},\frac{1}{2}\}$, where the mirror plane lies between two stacks. This symmetry allows for bond length variations within each individual stack, leading to distinct vertical bond lengths across the three stacks in the Three-UC slab, whereas only one stack exhibits such variations in the One-UC slab. This symmetry distinction is further reflected in the hopping parameters, influencing the electronic properties of the system.

For $U_{eff} = 2$ eV, electron correlation significantly influences the outer-apical Ni-O bond lengths ($d_1$ and $d_4$) in the Three-UC and One-UC slabs, whereas its effect is negligible in the Half-UC slab. Meanwhile, the in-plane Ni-O bond length ($d_5$) remains relatively stable in all configurations. These structural modifications are expected to have a pronounced impact on the electronic structures, which will be examined in detail in the following subsection. These findings highlight the critical role of interlayer interactions in determining the structural properties, which cannot be adequately captured by the Half-UC slab.

\subsection{Electronic structures}

\begin{figure*}
        \centering
	\includegraphics[scale=0.57]{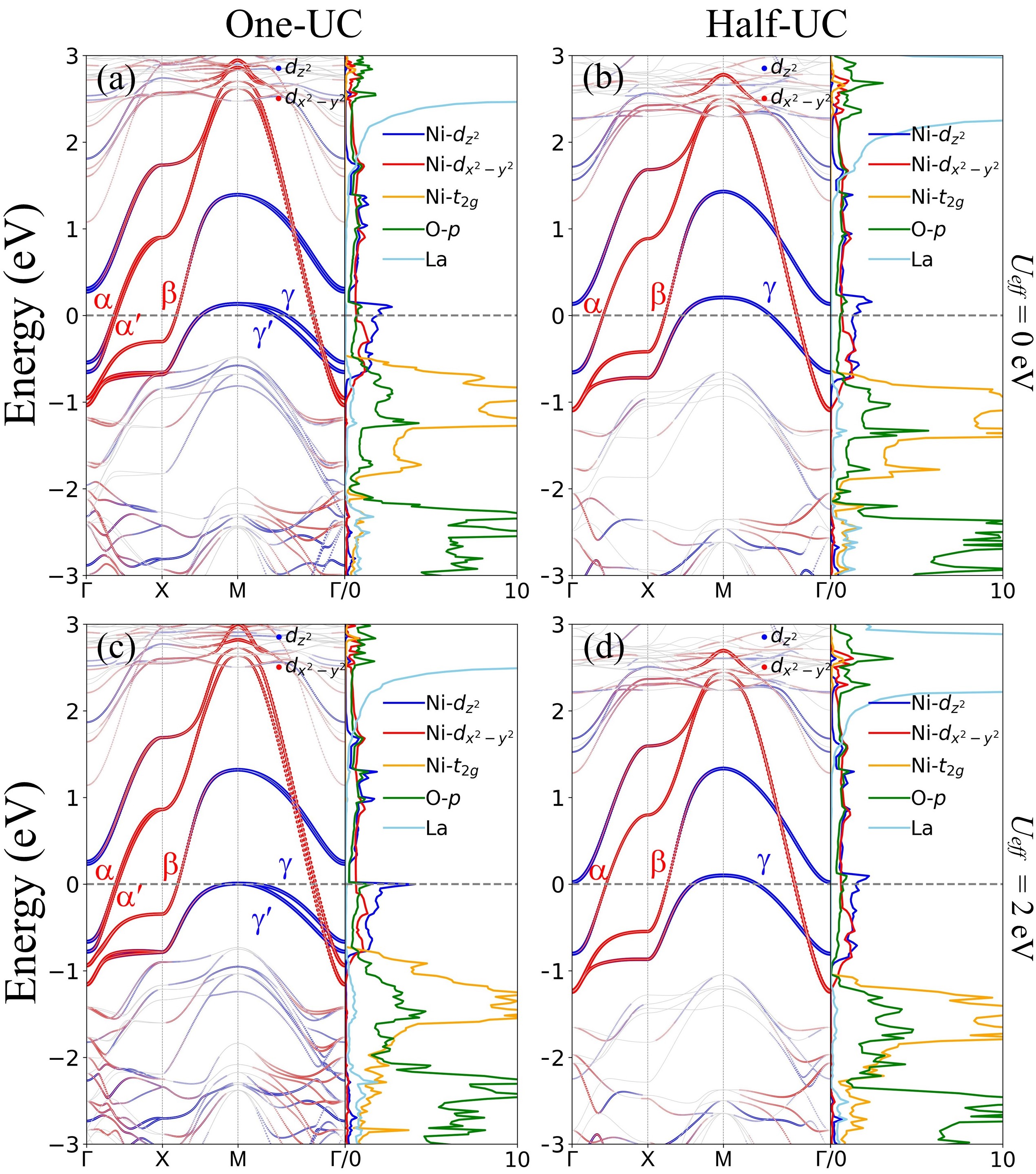}\caption{\label{fig2}DFT-calculated band structures and partial density of states (PDOS) of La$_3$Ni$_2$O$_7$ thin films. Panels (a) and (b) correspond to calculations with $U_{eff}=0$ eV, and panels (c) and (d) correspond to $U_{eff}=2$ eV. The contributions from the Ni$-d_{x^{2}-y^{2}}$ and $d_{z^2}$ orbitals are highlighted in blue and red, respectively, while Ni$-t_{2g}$, O-$p$, and La states are represented in orange, green, and cyan, respectively. The Fermi level ($E_{F}$) is set to 0 eV.
    Here, $U_{eff}$ denotes the effective hubbard parameter, and UC represents unit cell.}
\end{figure*}

\begin{figure}
        \centering
	\includegraphics[scale=0.75]{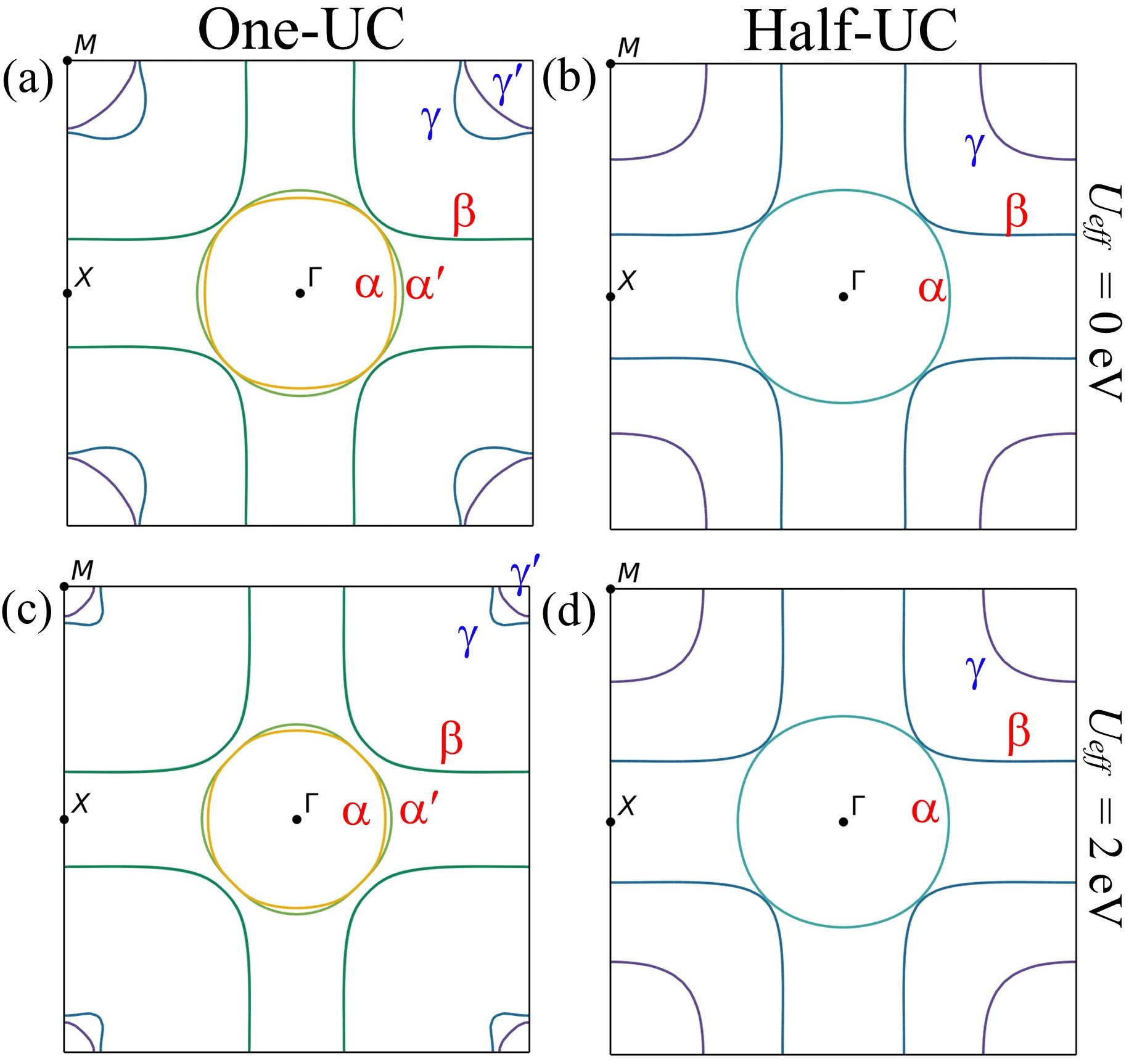}\caption{\label{fig3}DFT-calculated two-dimensional Fermi surfaces of La$_3$Ni$_2$O$_7$ thin films for different slab models. Panels (a) and (c) correspond to the One-UC structure, while panels (b) and (d) corresponds to the Half-UC structure. The upper panels show results obtained with $U_{eff}=0$ eV, and the lower panels correspond to $U_{eff} = 2$ eV. Here, $U_{eff}$ denotes the effective hubbard parameter, and UC represents unit cell.}
\end{figure}

We now investigate the electronic structures of La$_3$Ni$_2$O$_7$ thin films based on DFT calculations. For $U_{eff}=0$ eV, the band structure and projected density of states (PDOS) of the One-UC slab exhibit a clear metallic character, as shown in Figure~\ref{fig2}(a). The electronic states near the Fermi level ($E_{F}$) are primarily composed of Ni$-d_{x^2-y^2}$ and Ni$-d_{z^2}$ orbitals, which are well separated from the lower-energy Ni-$t_{2g}$ orbitals. Moreover, these Ni$-d$ orbitals exhibit hybridization with O$-p$ orbitals within the energy range of $-2$ eV to $2$ eV. Due to interlayer hybridization, the Ni$-d_{z^2}$ electronic states form bonding and antibonding bands, located below and above $E_{F}$, respectively. Additionally, the La-derived states make minimal contributions to the electronic states at $E_{F}$. The electronic structures of the Three-UC slab exhibit a similar behavior to that of the One-UC slab, as shown in Supplementary Figure 1(a). 
Notably, the reduced structural symmetry in thin films induces a splitting of Ni-$d_{z^2}$ bonding bands along the $M-\Gamma$ direction, leading to the formation of distinct electronic pockets near the near the $M$ point [See Figure \ref{fig3}(a)]. In contrast, the Half-UC slab, which retains higher structural symmetry, does not exhibit such splitting in the Ni-$d_{z^2}$ bonding bands [see Figure \ref{fig2}(b)]. 

For $U_{eff}=2$ eV, the Ni$-d_{z^2}$ bonding bands shift downward and approach $E_{F}$. Consequently, the density of states of Ni$-d_{z^2}$ attains a maximum at $E_{F}$ in the One-UC slab, as shown in Figure \ref{fig2}(c). A similar downward shift is observed in the electronic structure of the Three-UC slab, where the Ni$-d_{z^2}$ bonding bands move closer to $E_{F}$, as demonstrated in Supplementary Figure 1 (b). 
This behavior corresponds to the metallization of the lower $\sigma$ bonds in the high-presure phase of bulk La$_3$Ni$_2$O$_7$. These results indicate that the electronic structure of La$_3$Ni$_2$O$_7$ thin films closely resembles that of the high-pressure phase of the bulk material, which may provide insight into the emergence of high-temperature superconductivity in thin films. In contrast, the $E_{F}$ in the Half-UC slab exhibits only a slight downward shift, as illustrated in Figure \ref{fig2}(d). Previous studies on bulk La$_3$Ni$_2$O$_7$ have reported that DFT calculations with $U_{eff}=3.5$ eV yield results that are in good agreement with experimental observations~\cite{yang_orbital-dependent_2024}. To further explore this, we also consider the case of $U_{eff}=3.5$ eV. Under this condition, the Ni$-d_{z^2}$ bonding bands in the Three-UC and One-UC slabs shift further downward, moving below $E_{F}$, as shown in Supplementary Figures. 1(c) and 2(a). 
However, in the Half-UC slab, $E_{F}$ still crosses the Ni$-d_{z^2}$ bonding bands, as shown in Supplementary Figure 2(b). 

Figure~\ref{fig3} presents the evolution of the Fermi surface in La$_3$Ni$_2$O$_7$ thin films under different slab configurations and Hubbard $U_{eff}$. Panels (a) and (c) correspond to $U_{eff}=0$ eV, while panels (b) and (d) correspond to $U_{eff}=2$ eV. For $U_{eff}=0$ eV, the Fermi surface of the One-UC slab consists of three electron pockets ($\alpha,\alpha^{\prime},\beta$) and two hole pockets ($\gamma,\gamma^{\prime}$), as shown in Figure~\ref{fig3}(a). Notably, the $\gamma$ and $\gamma^{\prime}$ pockets are primarily derived from the Ni$-d_{z^2}$ orbital. The splitting of the Ni-$d_{z^2}$ and Ni$-d_{x^2-y^2}$ states along the $M-\Gamma$ and $X-\Gamma$ directions gives rise to the additional $\alpha^{\prime}$ and $\gamma^{\prime}$ pockets. This splitting originates from inter-stack interactions induced by the reduced structural symmetry in thin films. In contrast, there are two electron pockets ($\alpha,\beta$) and one hole pocket ($\gamma$) on the Fermi surface in the Half-UC slab, as shown in Figure~\ref{fig3}(b). 

For $U_{eff}=2$ eV, as the Ni$-d_{z^2}$ bonding bands in the One-UC slab shift downward and approach $E_{F}$, there is a significant reduction in the spatial extent of the hole pokets $\gamma$ and $\gamma^{\prime}$ around the $M$ point in the Brillouin zone, as shown in Figure~\ref{fig3}(c). In contrast, the Fermi surface geometry in the Half-UC slab exhibits minimal changes, as depicted in Figure~\ref{fig3}(d).


\subsection{Two-orbital models}
\begin{table}[t]
	\caption{\label{tab2} Hopping parameters for the One-UC slab model of La$_{3}$Ni$_{2}$O$_{7}$ thin films. $t_{A/B,[00]}^{x}$ and $t_{A/B,[00]}^{z}$ represent the site energies for the $d_{x^{2}-y^{2}}$ and $d_{z^2}$ orbitals in layer A and B, repectively. The numbers in the brackets correspond to parameters of the high-pressure phase of bulk La$_{3}$Ni$_{2}$O$_{7}$~\cite{luo2023bilayer}. The units are eV.} 
		\centering
		\begin{tabular}{ccccccc}
                \toprule
			\multicolumn{1}{c}{Index} & Layer & $i$ & $j$ & $t_{[ij]}^{x}$ & $t_{[ij]}^{z}$ & $t_{[ij]}^{xz}$ \\
			\hline
			\multirow{4}{*}{Stack 1}
			& A  & 0 & 0 &  0.844(0.776) &  0.519(0.409) & 0.000 \\ 
			&    & 1 & 0 & -0.462(-0.483) & -0.134(-0.110) & 0.228(0.239) \\
			&    & 1 & 1 &  0.075(0.069) & -0.021(-0.017) & 0.000 \\
			&    & 2 & 0 & -0.053 & -0.008 & 0.017 \\
			&    & 3 & 0 & -0.013 & -0.003 & 0.000 \\
			\cmidrule{2-7}
			& B  & 0 & 0 &  0.918 &  0.344 & 0.000 \\
			&    & 1 & 0 & -0.460 & -0.083 & 0.201 \\
			&    & 1 & 1 &  0.075 & -0.015 & 0.000 \\
			&    & 2 & 0 & -0.055 & -0.013 & 0.023 \\
			&    & 3 & 0 & -0.012 & -0.002 & 0.000 \\
			\cmidrule{2-7}
			& AB& 0 & 0 &  0.005(0.005)  & -0.550(-0.635) &  0.000\\
			&   & 1 & 0 & -0.000  &  0.020 & -0.031(-0.034)\\
			\hline
			\multirow{4}{*}{Stack 2} 
			& A  & 0 & 0 &  0.918 &  0.344 & 0.000 \\
			&    & 1 & 0 & -0.460 & -0.083 & 0.201 \\
			&    & 1 & 1 &  0.075 & -0.015 & 0.000 \\
			&    & 2 & 0 & -0.055 & -0.013 & 0.023 \\
			&    & 3 & 0 & -0.012 & -0.002 & 0.000 \\
			\cmidrule{2-7}
			& B  & 0 & 0 &  0.844 &  0.519  & 0.000 \\
			&    & 1 & 0 & -0.462 & -0.134 & 0.228 \\
			&    & 1 & 1 &  0.075 & -0.021 & 0.000 \\
			&    & 2 & 0 & -0.053 & -0.008 & 0.017 \\
			&    & 3 & 0 & -0.013 & -0.003 & 0.000 \\
			\cmidrule{2-7}
			& AB & 0 & 0 &  0.005 & -0.550 &  0.000\\
			&    & 1 & 0 &  0.000&  0.020 & -0.029\\
                \botrule
		\end{tabular}
\end{table}

\begin{table}[t]
	\caption{\label{tab3} Hopping parameters for Half-UC slab model of La$_{3}$Ni$_{2}$O$_{7}$ thin films. $t_{A,[00]}^{x}$, $t_{A,[00]}^{z}$ are the site energies for $d_{x^{2}-y^{2}}$ and $d_{z^2}$ orbitals in layer A. The units are eV.} 
		\centering
		\begin{tabular}{ccccccc}
                \toprule
			\multicolumn{1}{c}{Index} & Layer & $i$ & $j$ & $t_{[ij]}^{x}$ & $t_{[ij]}^{z}$ & $t_{[ij]}^{xz}$ \\
			\hline
			\multirow{4}{*}{Stack 1}
			& A  & 0 & 0 &  0.756 &  0.389 & 0.000 \\ 
			&    & 1 & 0 & -0.445 & -0.131 & 0.221 \\
			&    & 1 & 1 &  0.060 & -0.015 & 0.000 \\
			&    & 2 & 0 & -0.057 & -0.011 & 0.019 \\
			&    & 3 & 0 & -0.009 & -0.004 & 0.000 \\
			\cmidrule{2-7}
			& AB& 0 & 0 &  0.000 & -0.503 &  0.000 \\
			&   & 1 & 0 &  0.000 &  0.026 & -0.031 \\
                \botrule
		\end{tabular}
\end{table}
Based on the previous subsections, the thickness of the slab model significantly influences the Ni-O bond lengths, which in turn affect the electronic structure. Experimental samples typically have a thickness of One-UC--Three-UC, and calculations for the Three-UC slab are expected to more closely reflect the experimental conditions. However, due to the complexity of the Three-UC slab model, we have chosen to construct the tight-binding (TB) model using the One-UC slab, which captures the main features of the electronic band structure observed in the Three-UC slab. The Half-UC slab is also considered for comparison. 

Building on the electronic structures obtained from our DFT calculations, we first focus on the double-stacked two-orbital model for the One-UC slab. This model incorporates the Ni$-d_{x^{2}-y^{2}}$ and Ni$-d_{z^2}$ orbitals within the framework of the maximally localized Wannier functions Hamiltonian. The total Hamiltonian is given by
\begin{align}
\label{eq:H} 
\mathcal{H}&= \mathcal{H}_{0}+\mathcal{H}_U,  \\
\mathcal{H}_{0}&=\sum_{{\rm k}\sigma}\Psi_{{\rm k}\sigma}^{\dagger}H({\rm k})\Psi_{{\rm k}\sigma}. \nonumber
\end{align}
Here $\mathcal{H}_0$ denotes the TB Hamiltonian derived from the Wannier downfolding procedure, while $\mathcal{H}_U$ represents the Coulomb interaction term~\cite{kanamori}. 

The basis of the model is defined as $\Psi_{\sigma}=\left(d_{1Ax\sigma},d_{1Az\sigma},d_{1Bx\sigma},d_{1Bz\sigma},\\ 
d_{2Ax\sigma},d_{2Az\sigma},d_{2Bx\sigma},d_{2Bz\sigma}\right)^{T}$,
where the field operator $d_{s\sigma}$ annihilates an electron in the state $s$ with spin $\sigma$. The indices are assigned as follows: $1/2$ label the stacked layers, $A/B$ correspond to the bilayer sublattices, and $x/z$ denote the $d_{x^2-y^2}$ and $d_{z^2}$ orbitals, respectively. The labeling convention is illustrated in Figure~\ref{fig1}(b).

The TB Hamiltonian $H(\rm{k})$ takes the form 
\begin{align}	
H  ({\rm k})&=\left(\begin{array}{cc}
	H^{1}({\rm k}) & H^{12}({\rm k})\\
	H^{12}({\rm k}) & H^{2}({\rm k})
\end{array}\right),\nonumber \\
H^{1/2}  ({\rm k})&=\left(\begin{array}{cc}
H_{A}^{1/2}({\rm k}) & H_{AB}^{1/2}({\rm k})\\
H_{AB}^{1/2}({\rm k}) & H_{B}^{1/2}({\rm k})
\end{array}\right),\nonumber \\
H_{A/B}^{1/2}({\rm k})&=  \left(\begin{array}{cc}
H_{A/B}^{1/2,x}({\rm k}) & H_{A/B}^{1/2,xz}({\rm k})\\
H_{A/B}^{1/2,xz}({\rm k}) & H_{A/B}^{1/2,z}({\rm k})
\end{array}\right),\nonumber\\
H_{AB}^{1/2}({\rm k})&=\left(\begin{array}{cc}
H_{AB}^{1/2,x}({\rm k}) & H_{AB}^{1/2,xz}({\rm k})\\
H_{AB}^{1/2,xz}({\rm k}) & H_{AB}^{1/2,z}({\rm k})
\end{array}\right).\label{eq:tb}
\end{align}

The matrix elements are defined as follows:
\begin{align*}
H_{A/B}^{1/2,x/z}({\rm k})=
&2t_{A/B,[1,0]}^{1/2,x/z}\left(\cos  {k}_{x}+\cos {k}_{y}\right)\\+&2t_{A/B,[2,0]}^{1/2,x/z}\left(\cos  2{k}_{x}+\cos 2{k}_{y}\right)\\+&2t_{A/B,[3,0]}^{1/2,x/z}\left(\cos  3{k}_{x}+\cos 3{k}_{y}\right)\\+&4t_{A/B,[1,1]}^{1/2,x/z}\cos {k}_{x}\cos {k}_{y}+\epsilon_{A/B}^{1/2,x/z},\\
H_{A/B}^{1/2,xz}(\rm{ k})=&2t_{A/B,[1,0]}^{1/2,xz}\left(\cos {k}_{x}-\cos {k}_{y}\right)\\+&2t_{A/B,[2,0]}^{1/2,xz}\left(\cos 2{k}_{x}-\cos 2{k}_{y}\right),\\
H_{AB}^{1/2,x/z}(\rm{k})=&t_{AB,[0,0]}^{1/2,x/z}+2t_{AB,[1,0]}^{1/2,x/z}\left(\cos {k}_{x}+\cos {k}_{y}\right),\\ 
H_{AB}^{1/2,xz}(\rm{ k})=&2t_{AB,[1,0]}^{1/2,xz}\left(\cos {k}_{x}-\cos {k}_{y}\right),\\
H^{12,z}_{AB}({\rm k}) =& 4 t^{12,z}_{AB}[\cos ({k}_{x}/2)\cos ({k}_{y}/2)].
\end{align*}
Here $H_{A/B}^{1/2,x/z}({\rm k})$ and $H_{AB}^{1/2,x/z}(\rm{k})$ describe intralayer and interlayer hopping within the same orbitals ($d_{x^{2}-y^{2}}$ or $d_{z^2}$), respectively, while $H_{A/B}^{1/2,xz}(\rm{ k})$ and $H_{AB}^{1/2,xz}(\rm{ k})$ represent intralayer and interlayer hybridization between $d_{x^{2}-y^{2}}$ and $d_{z^2}$ orbitals. Additionally, $H^{12,z}_{AB}({\rm k})=4 t^{12,z}_{AB}\cos ({k}_{x}/2)\cos ({k}_{y}/2)$ describes inter-stack hopping within the $d_{z^2}$ orbital, with a corresponding hopping parameter $t^{12,z}_{AB}=-0.025$. Hopping parameters for the One-UC slab model of La$_{3}$Ni$_{2}$O$_{7}$ thin film are summarized in Table \ref{tab2}.

For the single-stacked two-orbital model of the Half-UC slab, the system exhibits layer symmetry and the absence of inter-stack coupling, as characterized by the conditions $H^{1}({\rm k})=H^{2}({\rm k})$, $H^{12}({\rm k})=0$, and $H_{A}^{1}({\rm k})= H_{B}^{1}({\rm k})$. These conditions indicate that the Hamiltonians of the individual layers are identical, there is no direct coupling between the stacks, and the interfacial hopping parameters are equivalent for both sublattices. The corresponding hopping parameters for the Half-UC slab model of La$_{3}$Ni$_{2}$O$_{7}$ thin film are provided in Table \ref{tab3}.



\begin{figure}
        \centering
	\includegraphics[scale=0.34,trim=24 0 0 0]{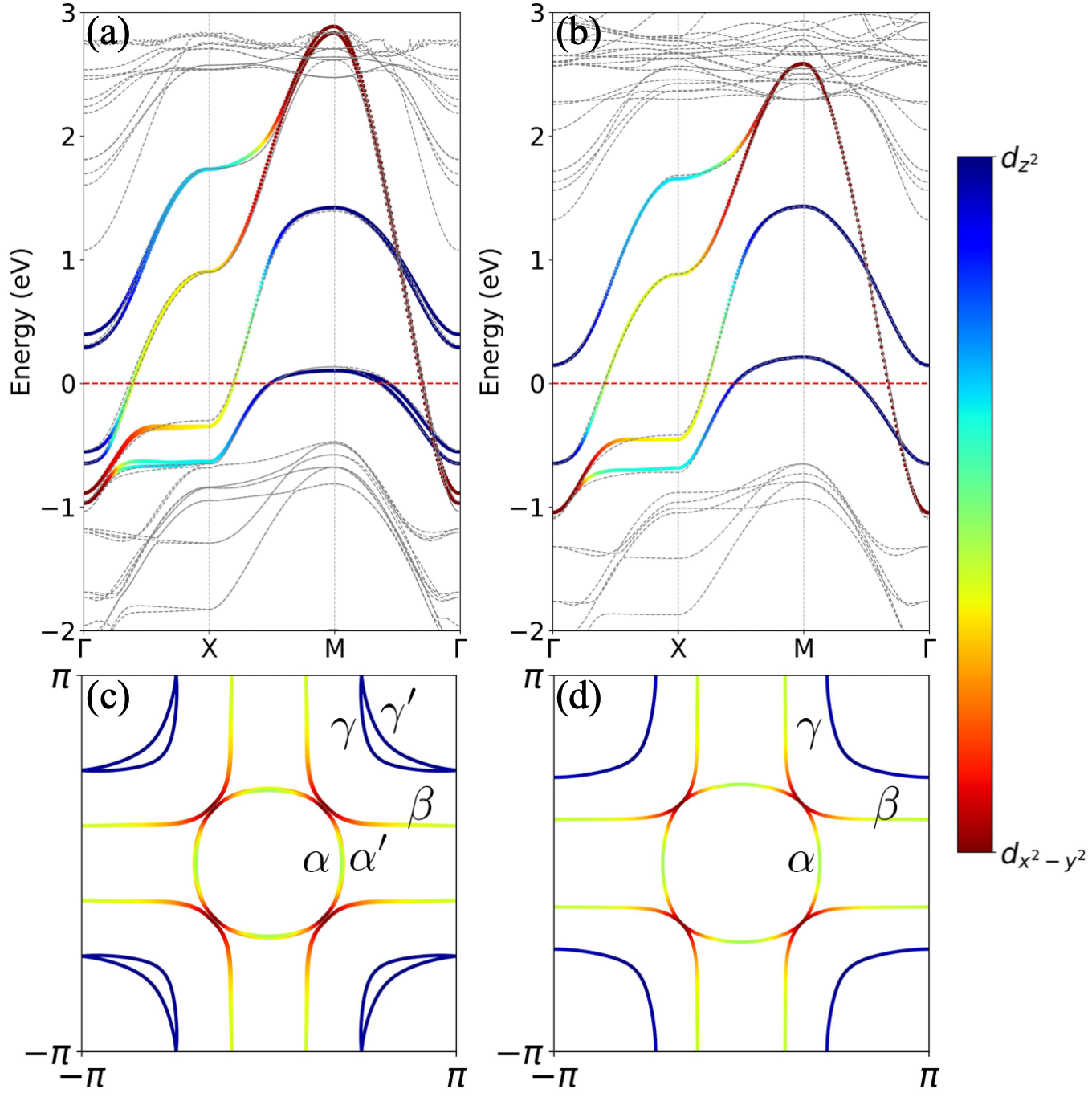}
	\caption{\label{fig4}Band structures and Fermi surfaces of the two-obital models for La$_3$Ni$_2$O$_7$ thin films. Panels (a) and (c) correspond to the double-stacked two-orbital model for the One-UC slab, while panels (b) and (d) correspond to the single-stacked two-orbital model for the Half-UC slab. The color bar indicates the orbital weights, with red and blue represent $d_{x^{2}-y^{2}}$ and $d_{z^2}$, respectively. In panels (a) and (b), the gray lines represent the DFT-calculated band structures with $U_{eff}=0$ eV. The $E_{F}$ is set to 0 eV. Here, $U_{eff}$ denotes the effective hubbard parameter, and UC represents unit cell.}
\end{figure}

Using the TB parameters listed in Tables \ref{tab2} and \ref{tab3}, we present the resulting band structure and Fermi surface for both One-UC and Half-UC slabs. As shown in Figure~\ref{fig4}(a), the model for the One-UC slab accurately reproduces the DFT band structure near the $E_\mathrm{F}$. Notably, we observe a splitting of the Ni-$d_{z^2}$ bonding bands along the $M-\Gamma$ direction, which originates from inter-stack interaction, specifically $t^{12,z}_{AB}$. Regarding the Fermi surface, we identify three electron pockets $\alpha,\alpha^{\prime},\beta$ and two hole pockets $\gamma,\gamma^{\prime}$, as illustrated in Figure~\ref{fig4}(c). The $\alpha,\alpha^{\prime},\beta-$pocket exhibit mixed orbital character, while the $\gamma,\gamma^{\prime}-$pockets are primarily dominated by the $d_{z^{2}}$ orbital state. Our results indicate that the electronic structure of La$_3$Ni$_2$O$_7$ thin film at ambient pressure closely resembles that of the high-pressure bulk phase. A key structural aspect is the in-plane lattice constant of the thin film, which is 3.77$\AA$--large than the pseudo-tetragonal bulk value of 3.715$\AA$. This lattice expansion may account for the lower $T_c$ of the thin film compared to the bulk, suggesting that strain engineering could be a viable approach to enhancing $T_c$ by further compressing the in-plane lattice constant. Furthermore, we note that the interlayer hopping amplitude $t_{AB,[00]}^{z}=-0.550$ is 1.19 times larger than the intralayer nearest-neighbor hopping $t_1^x=-0.462$. However, it remains 13.4$\%$ smaller than its bulk counterpart ($t_{AB,[00]}^{z}=-0.635$). This reduction in interlayer coupling suggests a possible weakening of unconventional pairing in the thin film compared to the bulk.

For the Half-UC slab model, no splitting in the Ni-$d_{z^2}$ bonding bands (See Figure~\ref{fig4}(b)). Consequently, the Fermi surface consists of  two electron pockets $\alpha,\beta$ and one hole pocket $\gamma$, as illustrated in Figure~\ref{fig4}(d). Additionally, the interlayer hopping amplitude $t_{AB,[00]}^{z}=-0.503$ is 1.13 times larger than the intralayer nearest-neighbor hopping $t_1^x=-0.445$, yet it remains 21$\%$ smaller than its bulk counterpart. Notably, when we set $t^{12,z}_{AB}=0$, no splitting is observed in the band structure or the Fermi surface of the One-UC slab, as shown in Supplementary Figure 3.

\subsection{High-energy $dp$ models}
\begin{table}[t]
    \centering
    \caption{\label{tab4}TB parameters for Wannier downfolding of the
		twenty-two-orbital model. Only the parameters of Stack 1 are presented, due to the symmetry between Stack 1 and Stack 2. See the schematic in Supplementary Figure 4
        for further details. }
        \centering
        \renewcommand{\arraystretch}{1.5}
        \begin{tabular}{cccc}
                \toprule
                \multicolumn{4}{c}{Stack 1} \\
                \hline
                \hline
               Hopping &   $Ad_{z^2}-p_{z'}$  &   $Ad_{z^2}-p_{z}$ &   $Ad_{z^2}-Ap_{x/y}$ \\
             \cmidrule{2-4}
              &                       1.304                      &   -1.478                      &   0.692   \\
             \hline
             \hline
             Hopping &   $Bd_{z^2}-p_{z''}$  &   $Bd_{z^2}-p_{z}$ &   $Bd_{z^2}-Bp_{x/y}$ \\     
            \cmidrule{2-4}
             &                       -1.105                      &   1.435                      &   0.629   \\
            \hline
             \hline
            Hopping &   $Ap_x-Ap_y$  &   $Ap_{x/y}-p_z$ &   $Ap_{x/y}-p_{z'}$ \\    
            \cmidrule{2-4}
            &                       -0.560                      &   -0.435                      &   0.443   \\
            \hline
             \hline
            Hopping &   $Bp_x-Bp_y$  &   $Bp_{x/y}-p_z$ &   $Bp_{x/y}-p_{z''}$ \\ 
            \cmidrule{2-4}
            &                       -0.575                      &   0.428                      &   -0.380   \\
            \hline
             \hline
             Hopping &   $Ad_{x^2-y^2}-Ap_{x/y}$  &   $Bd_{x^2-y^2}-Bp_{x/y}$ &   \\
            \cmidrule{2-4}
            &                       $\pm$1.480                      &   $\pm$1.479                      &      \\
            \hline
             \hline
            Site &   $Ad_{z^2}/Bd_{z^2}$  &   $Ad_{x^2-y^2}/Bd_{x^2-y^2}$ &     $p_z$ \\
            \cmidrule{2-4}
            Energy   &                       -1.104/-1.036                      &   -1.001/-0.9223                      &     -4.078 \\
            \hline
             \hline
            Site &   $Ap_{x/y}$/$Bp_{x/y}$  &   $Ap_{z'}/Bp_{z''}$ &     \\
            \cmidrule{2-4}
            Energy&                       -4.669/-4.630                      &   -2.730/-3.224                      &      \\
            \botrule

        \end{tabular}
        
\end{table}
\begin{table}[t]
    \centering
    \caption{\label{tab5}TB parameters for the Wannier downfolding of the eleven-orbital model. Only the parameters of Layer A are presented, due to the symmetry between Layers A and B. See the schematic in Supplementary Figure 4 for further details.
    }
        \centering
        \renewcommand{\arraystretch}{1.5}
        \begin{tabular}{ccccc}
                \toprule
                 \multicolumn{5}{c}{Stack 1} \\
                \hline
                \hline
                Hopping &   $Ad_{z^2}-p_z$   &   $Ad_{z^2}-p_{z'}$   &   $Ad_{z^2}-Ap_{x/y}$   &   $Ad_{x^2-y^2}-Ap_{x/y}$\\
                                \cmidrule{2-5}
                                &   -1.423   &  1.296   &   0.697   &   $\pm$1.482  \\
                \hline
                \hline
                Hopping &   $Ap_x-Ap_y$   &   $Ap_{x/y}-p_z$   &   $Ap_{x/y}-p_{z'}$   &      \\
                            \cmidrule{2-5}
                            &  -0.589  &   0.483  &     -0.440    &    \\
                \hline
                \hline
                Site    &   $Ad_{z^2}$ &   $Ad_{x^2-y^2}$ &   $Ap_{x/y}$ &   $p_z$ \\
                            \cmidrule{2-5}
                energy  &   -1.198  &   -1.127  &   -4.790   &    -4.150  \\
                \hline
                \hline
                Site    &   $Ap_{z'/z''}$ &    &    &   \\
                            \cmidrule{2-5}
                energy  &   -2.927  &     &      &    \\
                \botrule
                        
        \end{tabular}
        
\end{table}

\begin{figure}[ht]
    \centering
    \includegraphics[scale=0.2,trim=24 0 0 0]{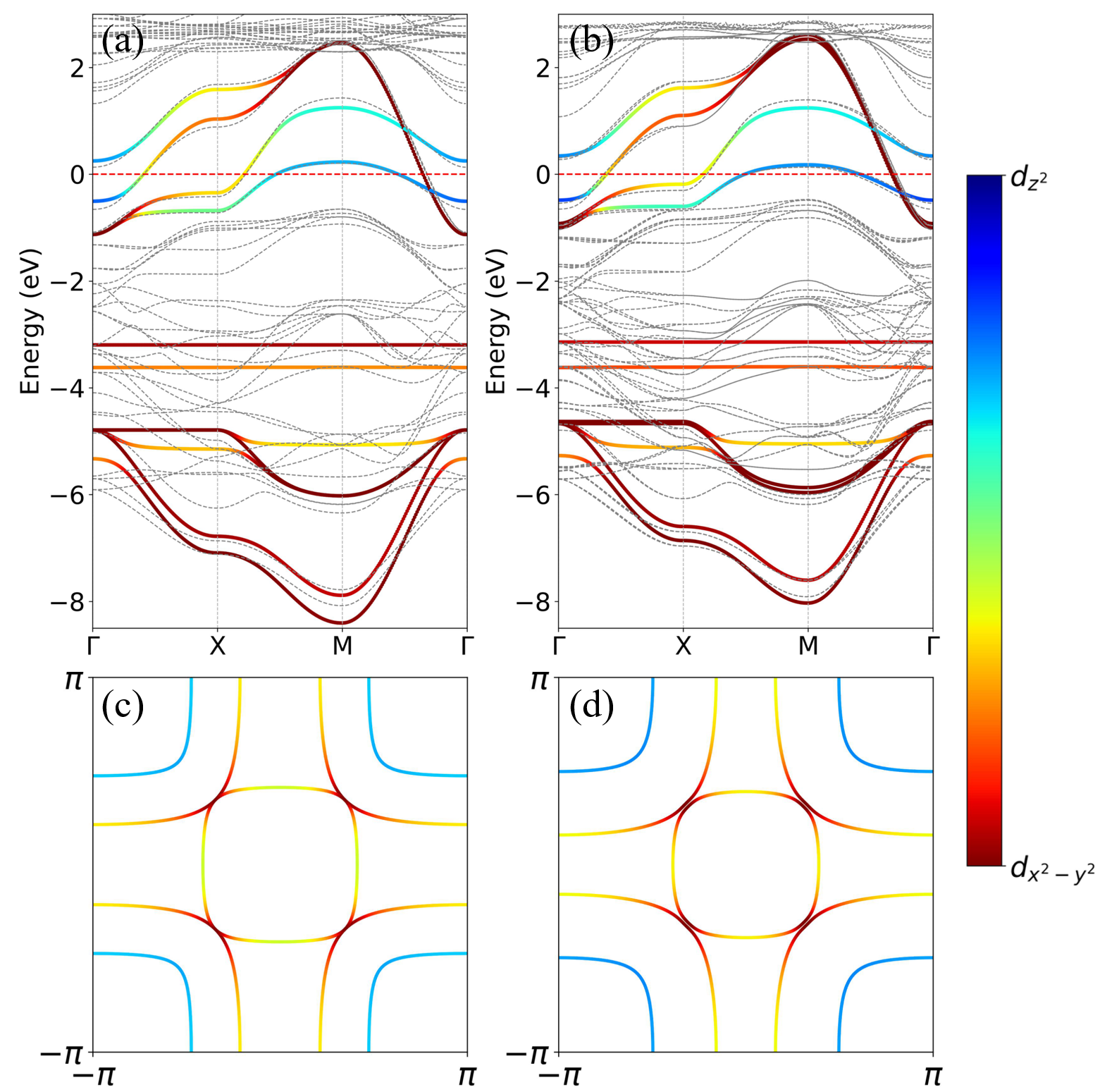}
    \caption{\label{fig5}Band structures and Fermi surfaces of High-energy $dp$ models including oxeygen orbitals for La$_3$Ni$_2$O$_7$ thin films. Panels (a) and (c) correspond to the eleven-orbital model for the Half-UC slab, while panels (b) and (d) correspond to the twenty-two-orbital model for the One-UC slab. The color bar indicates the orbital weights, with red and blue represent $d_{x^{2}-y^{2}}$ and $d_{z^2}$, respectively. In panels (a) and (b), the gray lines represent the DFT-calculated band structures with $U_{eff}=0$ eV. The $E_{F}$ is set to 0 eV. Here, $U_{eff}$ denotes the effective hubbard parameter, and UC represents unit cell.}
\end{figure}

To incorporate the effects of O$-p$ orbitals, we introduce high-energy $dp$ models: a double-stacked eleven-orbital model for the One-UC slab (referred to as twenty-two-orbital model) and a single-stacked eleven-orbital model for the Half-UC slab (referred to as the eleven-orbital model). In both models, the basis for Stack 1 is given by $\Psi=(d_{Az},d_{Bz},d_{Ax},d_{Bx},d_{Ap_x},d_{Bp_x},d_{Ap_y},d_{Bp_y},d_{p_z},d_{p_{z'}},d_{p_{z''}})^T$, which includes four in-plane orbitals ($p_{Ax},p_{Ay},p_{Bx},p_{By}$) and three apical orbitals ($p_z,p_{z'},p_{z''}$), as illustrated in Supplementary Figure 4. 
Here, $A$ and $B$ denote the bilayer sublattices.

The TB parameters of the twenty-two-orbital model are listed in Table~\ref{tab4}, while those of the eleven-orbital model are provided in Table~\ref{tab5}. Due to the symmetry between Stack 1 and Stack 2, only the parameters for Stack 1 are presented, and inter-stack hopping terms are not considered. Both models incorporate hopping interactions arising from $pd,pp$ orbital overlaps. Based on the TB parameters, we present the resulting band structure and Fermi surface for both the One-UC and Half-UC slabs, as shown in Figure~\ref{fig5}. 
The resulting band structure in Figure~\ref{fig5} covers an energy range akin to that of Figure~\ref{fig4} and can also reproduce the main features at $E_\mathrm{F}$. 
Furthermore, we find a hopping of 1.296 for eleven-orbital model and of 1.304/1.105 for twenty-two-orbital model between $d_{z^2}$ orbital and two apical $p_{z'},p_{z''}$ orbitals, which are crucial in estimating effective interlayer $d_{z^2}$ spin exchange coupling.   
The high-energy models would provide a foundation for further investigations of magnetic exchange coupling and electronic correlations. 

\subsection{Spin susceptibility for two-orbital models}
\begin{figure}
        \centering
	\includegraphics[scale=0.32,trim=24 0 0 0]{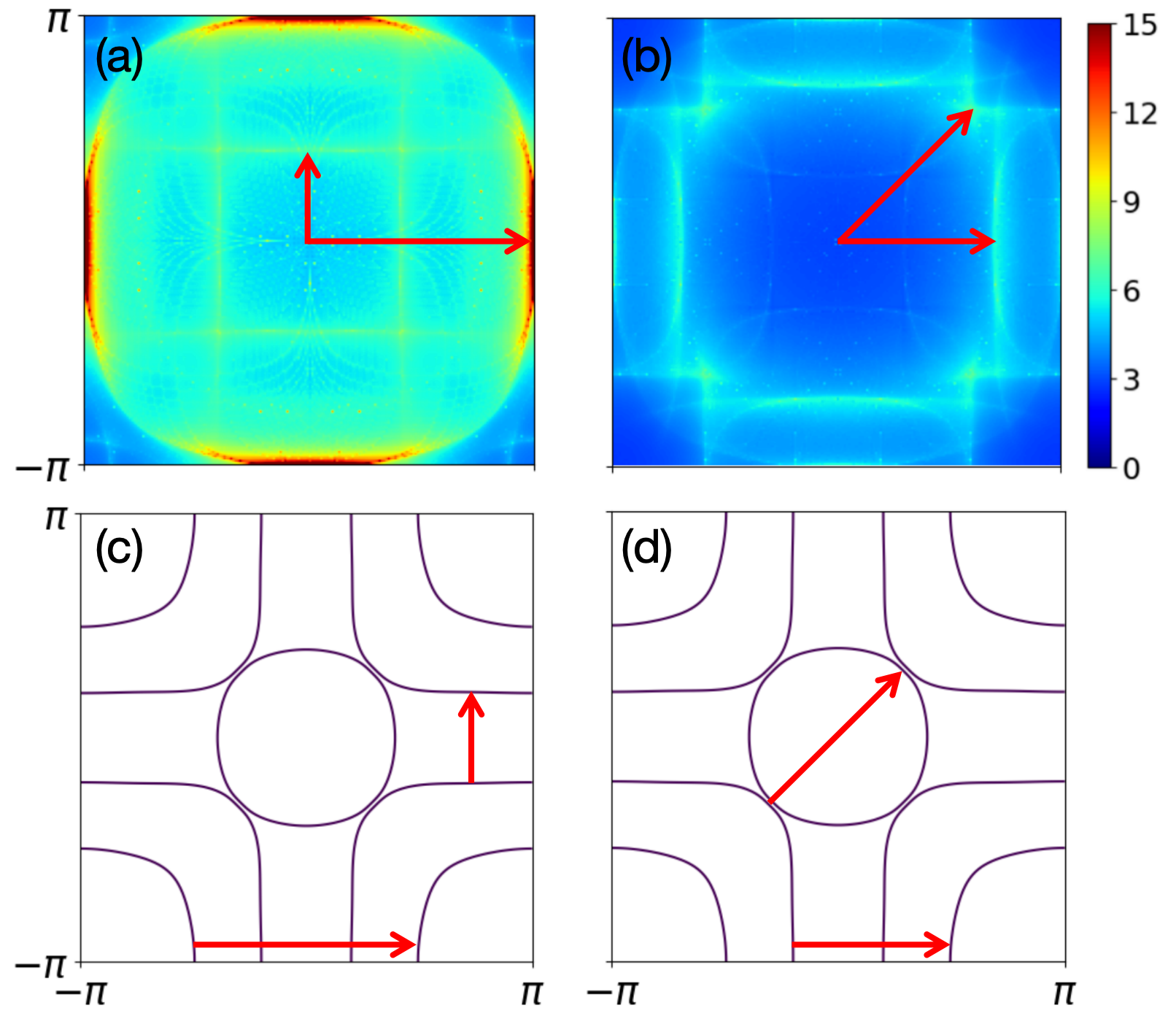}\caption{\label{fig7}
    RPA spin susceptibilities of even and odd channel, and illustration of Fermi surfaces nesting in the double-stacked two-orbital model.
    Panels (a) and (b) show the even channel (\(\chi^e\)) and odd channel (\(\chi^o\)) susceptibilities for the double-stacked two-orbital model, respectively, calculated with intraorbital Coulomb interaction \(U=0.7\) eV and Hund's coupling \(J_H=0.2\) eV at a temperature \(T=0.1 \ \mathrm{K}\). Panels (c) and (d) depict the Fermi surfaces, where red arrows indicating the nesting vectors.}
\end{figure}
\begin{figure}
        \centering
	\includegraphics[scale=0.275,trim=24 0 0 0]{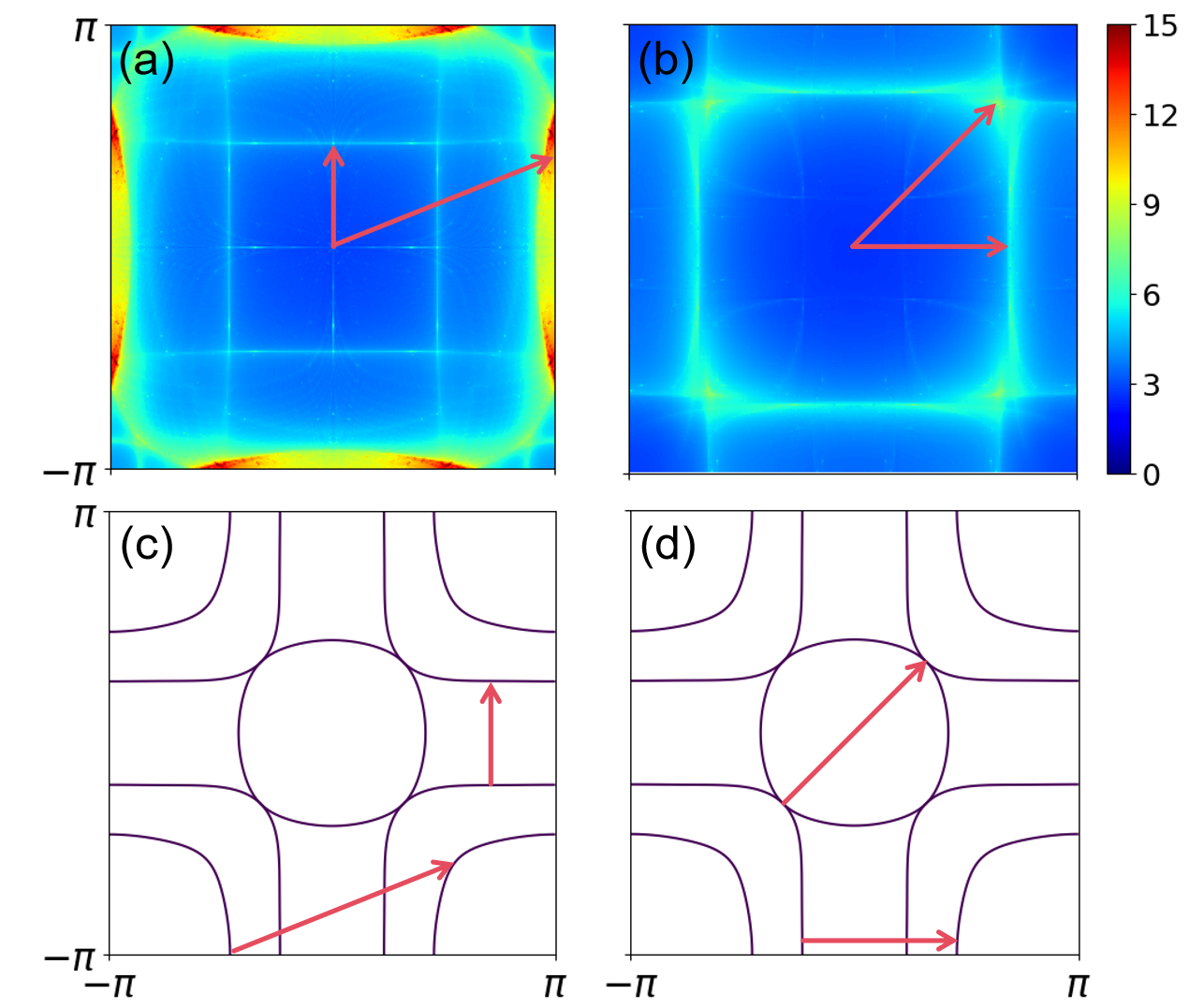}\caption{\label{fig8}
    RPA spin susceptibility of even and odd channel, and illustration of Fermi surfaces nesting in the single-stacked two-orbital model.
    Panels (a) and (b) show the even channel (\(\chi^e\)) and odd channel (\(\chi^o\)) susceptibilities for the single-stacked two-orbital model, respectively, calculated with intraorbital Coulomb interaction \(U=0.8 \) eV and Hund's coupling \(J_H=0.2\) eV at a temperature \(T=0.1 \ \mathrm{K}\). Panesls (c) and (d) present the Fermi surfaces, where red arrows indicating the nesting vectors.}
\end{figure}

With the multi-orbital Hubbard model defined above, we investigate the magnetic response and Fermi surface nesting by calculating the spin susceptibility at the RPA level. Given that inter-stack hoppings are weak, we neglect them in our analysis. The resulting Fermi surface and energy bands in the absence of inter-stack hopping are provided in the Supplementary Figure 3.

To better demonstrate the Fermi surface nesting relations associated with the bilayer structure, we define the even and odd magnetic susceptibilities as: \(\chi^{e/o} = \chi_{||} \pm \chi_{\bot}\),  with $\chi_{||}$ and $\chi_{\bot}$ representing intralayer and interlayer contributions, respectively.  The definitions and computational details are available in the Methods section.
Physically, it is easily proved that the $\chi^{e}$ originates from nestings within the bonding ($\alpha,\gamma$) and within the antibonding ($\beta$) bands, while $\chi^{o}$ originates solely from nestings between the bonding and antibonding bands~\cite{PhysRevB.109.L180502}.


Figure~\ref{fig7} presents the static RPA spin susceptibility $\chi^{e/o}(\mathbf{q},\omega=0)$ for the double-stacked two-orbital model, computed with \(U=0.7\) eV, \(J_H=0.2\) eV, and temperature \(T=0.1 \ \mathrm{K}\). In the even channel, we observe a strong response near the \(X\) point,  indicating nesting within the \(\gamma\) pocket, along with additional nesting within the \(\beta\) pocket. In the odd channel, nesting is evident between the \(\beta\) and \(\gamma\) pockets, as well as between the \(\alpha\) and \(\beta\) pockets.

Figure~\ref{fig8} presents the spin susceptibility for the single-stacked two-orbital model, calculated using \(U=0.8\) eV, \(J_H=0.2\) eV and \(T=0.1 \ \mathrm{K}\) . The Fermi surface nesting patterns in the odd channel closely resemble those of the double-stacked model. In the even channel, nesting primarily occurs within the \(\gamma\) and \(\beta\) pockets, whereas in the odd channel, significant nesting is observed between the \(\gamma\) and \(\beta\) pockets, as well as between the \(\alpha\) and \(\beta\) pockets. The observed differences in $\chi^{e/o}$ between the double-stacked and single-stacked models can be attributed to their distinct Fermi surface geometries.



\section{Methods}\label{sec3}

\subsection{First-principles calculations}

Our first-principles calculations are performed using the DFT as implemented in the Vienna {\it ab initio} simulation package (VASP)~\cite{93prb,96prb}. The exchange-correlation interactions are treated within the generalized gradient approximation (GGA) using the Perdew-Burke-Ernzerhof (PBE) functional~\cite{96prl}. The projector augmented-wave (PAW) method~\cite{94prb} is employed with a plane-wave cutoff energy of 600 eV. Structural relaxation and electronic self-consistent calculations are conducted on a $\Gamma$-centered $12\times 12\times 1$ $k-$points mesh using the Monkhorst-Pack scheme, while a denser $k-$points grid of $27\times 27\times 1$ is used for Fermi surface calculations. The atomic positions are fully relaxed until the residual forces on each atom are less than 0.005 eV/$\AA$, and the electronic self-consistency convergence criterion is set to $10^{-6} $ eV. The optimized atomic coordinates are provided in Supplementary Data 1 and 2 for the Half-UC structure with $U_{eff}$ = 0 and 2 eV, respectively, and in Supplementary Data 3 and 4 for One-UC structure with $U_{eff}$ = 0 and 2 eV, respectively.


To construct effective models, maximally localized Wannier functions are obtained using the Wannier90 code~\cite{2020jpcm,97prb,2001prb}. For DFT $+~U$ calculations~\cite{98prb}, an effective Hubbard $U_{eff}$ is applied to the Ni 3$d$ orbitals. 

\subsection{Hamiltonian of high-energy $dp$ models}

For the twenty-two-orbital model, we do not consider the hoppings between orbitals in different stacks. So the hamiltonian of twenty-two-orbital model take the same form as the one for the eleven-orbital model. Here we show the TB hamiltonian for high-energy $dp$ models~\cite{luo2023bilayer}. The basis here is $(Ad_z, Bd_z, Ad_x, Bd_x, Ap_x, Bp_x, Ap_y, Bp_y, p_z, p_{z^{\prime}}, p_{z^{\prime\prime}})$. The position of these orbitals can be seen in Supplementary Figure 4. 
Here we show the elements of Hamiltonian $H(\alpha,\beta)$
\begin{align}
    H(1,9)= t^{Ad_z, p_z}, H(2,9)=t^{Bd_z,p_{z}}, H(1,10)=t^{Ad_z,p_{z^{\prime}}},  H(2,11)=t^{Bd_z,p_{z^{\prime\prime}}}, \\ \notag
    H(3,5)=-2\mathrm{i}t^{Ad_x,Ap_x}\sin(0.5k_x), \quad H(3,7)=-2\mathrm{i}t^{Ad_x-Ap_y}\sin(0.5k_y),
    \\ \notag
    H(4,6)=-2\mathrm{i}t^{Bd_x,Bp_x}\sin(0.5k_x), \quad H(4,8)=-2\mathrm{i}t^{Bd_x-Bp_y}\sin(0.5k_y),
    \\ \notag
    H(1,5)=-2\mathrm{i}t^{Ad_z,Ap_x}\sin(0.5k_x), \quad H(1,7)=-2\mathrm{i}t^{Ad_z-Ap_y}\sin(0.5k_y),
    \\ \notag
    H(2,6)=-2\mathrm{i}t^{Bd_z,Bp_x}\sin(0.5k_x), \quad H(2,8)=-2\mathrm{i}t^{Bd_z-Bp_y}\sin(0.5k_y),
    \\ \notag
    H(5,7)=4t^{Ap_x-Ap_y}\sin(0.5k_x)\sin(0.5k_y), \\ \notag
    H(6,8)=4t^{Bp_x-Bp_y}\sin(0.5k_x)\sin(0.5k_y), \\ \notag
    H(5,9)=-2\mathrm{i}t^{Ap_x,p_z}\sin(0.5k_x), \quad H(7,9)=-2\mathrm{i}t^{Ap_y,p_z}\sin(0.5k_y),
    \\ \notag
    H(6,9)=-2\mathrm{i}t^{Bp_x,p_z}\sin(0.5k_x), \quad H(8,9)=-2\mathrm{i}t^{Bp_y,p_z}\sin(0.5k_y),
    \\ \notag
    H(5,10)=-2\mathrm{i}t^{Ap_x,p_{z^{\prime}}}\sin(0.5k_x),  \quad H(7,10)=-2\mathrm{i}t^{Ap_y,p_{z'}}\sin(0.5k_y),
    \\ \notag
    H(6,11)=-2\mathrm{i}t^{Bp_x,p_{z''}}\sin(0.5k_x), \quad H(8,11)=-2\mathrm{i}t^{Bp_y,p_{z''}}\sin(0.5k_y).\\ \notag
\end{align}
The elements $H(\beta,\alpha)$ can be obtained by $H(\beta,\alpha)=H(\alpha,\beta)^{*}$. The diagnal elements are site energies which can be found in Table~\ref{tab4} and Table~\ref{tab5}.

\subsection{Calculation of spin susceptibility}
Upon determining the hopping parameters of the TB models and incorporating electron interactions, we construct a multi-orbital Hubbard model as Equation~\ref{eq:H} without considering hoppings between two stacks.
\begin{align}
	\mathcal{H}_{U} =& U \sum_{is}n_{is\uparrow} n_{is\downarrow}
	  \\
	+& (U^{\prime}-J_H \delta_{\sigma\sigma^{\prime}}) 
	\sum_{i\sigma \sigma^{\prime}} (n_{iAx\sigma}n_{iAz\sigma^{\prime}}
	+n_{iBx\sigma}n_{iBz\sigma^{\prime}}) 
	\notag \\
	+& J_H\sum_{i\sigma} \sum_{\mu}^{A,B}
	(d^{\dagger}_{i\mu x\sigma} d^{\dagger}_{i\mu z\bar{\sigma}} d_{i\mu x\bar{\sigma}} d_{i\mu z\sigma}
	\notag \\
	+& d^{\dagger}_{i\mu x\sigma} d^{\dagger}_{i\mu x\bar{\sigma}} d_{i\mu z\bar{\sigma}} d_{i\mu z\sigma} + h.c.).  \notag
\end{align}
The TB Hamiltonian \(\mathcal{H}_0\) is expressed with the basis \(\Psi_{\sigma}=(d_{Ax\sigma}, d_{Az\sigma}, d_{Bx\sigma}, d_{Bz\sigma})\), where \(d_{s\sigma}\) is the annihilation operator for an electron in the state \(s=(Ax,Az,Bx,Bz)\) with spin \(\sigma\). The interaction parameters include
\(U\) (intraorbital Coulomb interaction), \(U^{\prime}\) (interorbital Coulomb interaction) and \(J_H\) (Hund's coupling). Kanamori relation is applied here, given by \(U^{\prime}=U-2J_H\) \cite{annurev:/content/journals/10.1146/annurev-conmatphys-020911-125045}.

In general, the bare (non-interaction) susceptibility is defined as 
\begin{align}
	\chi^{0}_{\alpha \beta \gamma \delta}(\mathrm{q},\omega)
	&= -\frac{1}{N_\mathrm{k}} \sum_{\mathrm{k},mn} 
	\frac{f_F(\varepsilon_{\mathrm{k}}^{m}) - f_F(\varepsilon_{\mathrm{k+q}}^n)}{\mathrm{i}\omega+\varepsilon_{\mathrm{k}}^m-\varepsilon_{\mathrm{k+q}}^n}
	\notag \\
	&U_{\delta m}(\mathrm{k}) U_{\alpha m}^*(\mathrm{k}) U_{\beta n}(\mathrm{k+q}) U_{\gamma n}^{*}(\mathrm{k+q}).
\end{align}
with the band indices \(m, n\) and 
the Fermi-Dirac distribution function \(f_F(\varepsilon_{\mathbf{k}})=1/(\mathrm{e}^{\varepsilon_{\mathbf{k}}/T}+1)\) . 
The matrix element \(U_{\delta m}(\mathrm{k})\) represents the eigenvector connecting orbital \(\delta\) and band \(m\) at wave vector \(\mathrm{k}\). At the RPA level, the spin channel interaction vertex is defined as
\begin{align}
	\Gamma^{m}_{\alpha \beta \gamma \delta} = 
	\begin{cases}U & \alpha=\beta=\gamma=\delta, \\ 
		U^{\prime} & \alpha=\delta \neq \beta=\gamma, \\ 
		J_H & \alpha=\beta \neq \gamma=\delta, \\ 
		J^{\prime} & \alpha=\gamma \neq \beta=\delta.
	\end{cases}
\end{align}
Here pair hopping \(J^{\prime}\) satisfies \(J^{\prime}=J_H\).
The RPA spin susceptibility is then computed in a matrix-product form as
\begin{align}
	\chi_{(\alpha \beta, \delta \gamma)}^{S}
	&=\left[I-\chi^0 \Gamma_{(\alpha \beta, \delta \gamma)}^m\right]^{-1} 
	\chi_{(\alpha \beta, \delta \gamma)}^0.
\end{align}

Considering the bilayer structure of La$_3$Ni$_2$O$_7$, the interlayer hopping terms in the Hamiltonian acquire a phase factor of $\mathrm{e}^{\mathrm{i}k_z}$ due to the $k_z-$dependence. Here, $k_z$  can take values of either 0 or $\pi$. When contracting the orbital indices to compute the spin susceptibility, we define the in-plane and interlayer susceptibilities as $\chi_{||}=\sum_{\alpha\beta}(\chi_{A\alpha A\beta}+\chi_{B\alpha B\beta})$ and $\chi_{\bot}=\sum_{\alpha\beta} (\chi_{A\alpha B\beta}+\chi_{B\alpha A\beta})$, where $\chi_{\alpha\beta}=\chi_{\alpha\alpha\beta\beta}$ with orbital indices contracted. For $k_z=0$, the phase factor $\mathrm{e}^{\mathrm{i}0}=1$, and the spin susceptibility is given by $\chi^e=\chi_{||}+\chi_{\bot}$. Conversely, for $k_z=\pi$, the phase factor $\mathrm{e}^{\mathrm{i}\pi}=-1$, leading to a spin susceptibility of $\chi^o=\chi_{||}-\chi_{\bot}$. In the even channel $\chi^e$, Fermi surface nesting which occurs within the bonding/antibonding bands is evident, including the $\alpha-\alpha$, $\alpha-\gamma$, $\gamma-\gamma$ and $\beta-\beta$ nesting features. In the odd channel $\chi^o$, nesting between bonding and antibonding bands can be observed clearly, such as $\gamma-\beta$, $\alpha-\beta$~\cite{PhysRevB.109.L180502}.

\section{Data Availability}
The data that support the findings of this study are available from the corresponding author upon reasonable request.

\section{Discussion}\label{sec4}

Our DFT calculations of the thin-film bilayer nickelate superconductors have overall predicted the electronic structure closely resembles the bulk one under pressure, which strongly suggests that they all within the same superconducting mechansim. We note another key aspect to validate this idea, which  is the superexchange strength. 
For the bulk samples, the vertical superexchange between two half-filling $d_{z^2}$ orbitals $J_{\bot}^z$ is widely believed to be the origin of the superconducting condensation~\cite{wu2024superexchange, luo2024high}.
Here our TB models for the La$_3$Ni$_2$O$_7$ thin-film also allow an estimation of the corresponding strengths, which are decreased by $\sim$24\% and $\sim$36\% for single-stacked and double-stacked bilayer nickelates with respect to the pressurized bulk~\cite{luo2023bilayer}, assuming $J_\bot^z\sim (t^z)^2$.
The asymptotic decrease of $J_{\bot}^z$ with slab number from our estimation is in line with the notable decrease of $T_c$ from bulk to film as observated in experiments~\cite{ko_signatures_2025,zhou_ambient-pressure_2025}, which again reinforce the above general understanding. More profoundly, the consistency indicates an expension of the available experimental techniques  for reaching the core of  superconducting mechanism in RP nickelates. This is particularly helpful as the exploration on bulk  is largely constrained by the exerted pressure. Regrading the role of pressure, on the other hand, the existing evidences on thin-film is also prone to the idea that pressure can help suppressing spin density wave, after which superconducting order is thus to expose. Whereas for the thin-film, due to the higher stability and sniffiness of the  $\mathrm{SrLaAlO_4}$ substrates, $\mathrm{La_3Ni_2O_7}$ lattice is strongly confined to the $I4/mmm$ symmetry without  inplane distrostion, which can further prevent the occurance of density waves~\cite{ko_signatures_2025,zhou_ambient-pressure_2025}. 

On the other hand, it is worth noting that the structural differences between the One-UC and Half-UC slab models--particularly the presence of inter-stack geometry in the One-UC case--lead to more three-dimensional electronic characteristics. In particular, the enhanced interlayer coupling in the One-UC structure gives rise to noticeable out-of-plane ($k_z$) direction. This suggests that such $k_z$-dependent features may be experimentally observable in future high-resolution ARPES measurements, providing an additional means to probe the dimensionality and electronic reconstruction in La$_3$Ni$_2$O$_7$ thin films.

We note that both experimental and theoretical studies have reported differing results regarding the Fermi surface topology. While Ref.\cite{wang2025electronicstructurecompressivelystrained} reports a lowering of the Ni-d$_{z^2}$ states under compressive strain, an independent group has observed a d$_{z^2}$-derived $\gamma$-pocket at the Fermi level in La$_2$PrNi$_2$O$_7$ thin films, accompanied by a measurable superconducting gap on this pocket~\cite{li2025photoemissionevidencemultiorbitalholedoping,shen2025nodelesssuperconductinggapelectronboson}. These contrasting observations suggest that the presence or absence of the $\gamma$-pocket may be sample-dependent, potentially influenced by factors such as oxygen stoichiometry or subtle strain variations. On the theoretical side, previous studies~\cite{PhysRevB.111.115154,bhatta2025structuralelectronicevolutionbilayer} examined the effect of biaxial strain on bulk nickelates and found that compressive strain tends to increase the apical Ni–O–Ni bond angle toward 180$^{\circ}$, consistent with structural features of the high-pressure phase. Other works~\cite{geisler2025fermisurfacereconstructionenhanced,geisler2025electronicreconstructioninterfaceengineering}, adopting the lower-pressure Amam phase as a reference and employing DFT $+~U$ methods, reported the emergence of d$_{z^2}$-derived pockets at the Fermi surface under tensile strain. It is well established that the calculated electronic structure of La$_3$Ni$_2$O$_7$ is highly sensitive to the choice of the Hubbard $U_{eff}$ parameter (see Supplementary Figure 1). 
We anticipate that future experimental studies will help resolve these discrepancies and clarify the microscopic origin of the observed differences.


\section{Conclusion}\label{sec5}

In conclusion, we employ slab models for La$_3$Ni$_2$O$_7$ thin films that simulate the electronic structure for various thicknesses, including Three-UC, One-UC, and Half-UC. Each slab model incorporates a full unit cell, with the two bilayers referred to as Stack 1 and Stack 2, enabling a detailed examination of the interplay between dimensionality and electronic behaviors. Using density functional theory, we propose a double-stacked two-orbital effective model of La$_3$Ni$_2$O$_7$ thin films, based on the Ni$-e_g$ orbitals. Our analysis reveals the presence of three electron pockets $\alpha,\alpha’,\beta$ and two hole pockets $\gamma,\gamma'$ on the Fermi surface, where the additional pockets $\alpha’$ and $\gamma’$ emerge due to inter-stack interactions. Furthermore, we introduce high-energy models incorporating O$-p$ orbitals to facilitate future studies. Spin susceptibility calculations within the RPA indicate pronounced magnetic correlations primarily driven by nesting effects of the $\gamma$ pocket, which is predominantly contributed by the Ni$-d_{z^2}$ orbital state. Our results provide theoretical framework for understanding the interplay among dimensionality, magnetism, and superconductivity in  La$_3$Ni$_2$O$_7$ thin films, offering key insights for future theoretical and experimental research.

\backmatter






\bibliography{sn-bibliography}
\bmhead{Acknowledgements}

We are grateful to W\'ei W\'u and Xiao-Hong Pan for useful discussions. Work at Sun Yat-Sen University was supported by the National Natural Science Foundation of China (Grants No. 12494591, No. 92165204), the National Key Research and Development Program of China (Grant No. 2022YFA1402802), Guangdong Provincial Key Laboratory of Magnetoelectric Physics and Devices (Grant No. 2022B1212010008), Research Center for Magnetoelectric Physics of Guangdong Province (Grant No. 2024B0303390001), Guangdong Provincial Quantum Science Strategic Initiative (Grant No. GDZX2401010), and Leading Talent Program of Guangdong Special Projects (201626003). 

\bmhead{Author contributions}
D.X.Y. and X.H. proposed and designed the project. X.H. and C.Q.C. contributed to DFT calculations. X.H. and D.X.Y. contributed to two-orbital models. W.Q. and D.X.Y. contributed to high-energy models. W.Q. and Z.L. contributed to spin susceptibility and discussion under the supervision of D.X.Y. All authors contributed to the interpretation of the results and wrote the paper.

\bmhead{Competing interests}
The authors declare no competing interests.

\bmhead{Materials and Correspondence}
Correspondence and requests for materials should be addressed to D.X.Y.

\end{document}